\documentclass[reprint,aps,prl,longbibliography,footinbib,floatfix,nobalancelastpage]{revtex4-2}
\usepackage{amsmath}
\usepackage{amsfonts}
\usepackage{amssymb}
\usepackage{mathtools}
\usepackage{graphicx}
\usepackage[dvipsnames]{xcolor}
\usepackage[colorlinks=True,citecolor=gray,linkcolor=myRed,urlcolor=gray,hypertexnames=false]{hyperref}
\usepackage{hypcap}
\usepackage{yfonts}
\usepackage{bm}
\usepackage[normalem]{ulem}
\usepackage{dsfont}
\usepackage{physics}
\usepackage{marginnote}

\newcommand\eq[1]{\begin{align}#1\end{align}}

\newcommand\sr[1]{{\color{black}{#1}}}

\definecolor{myBlue}{RGB}{31,119,180}
\definecolor{myOrange}{RGB}{255,127,14}
\definecolor{myGreen}{RGB}{44,160,44}
\definecolor{myRed}{RGB}{214,39,40}
\definecolor{myPurple}{RGB}{148,103,189}

\makeatletter
\def\p@figure{\color{myBlue}}
\def\p@equation{\color{myRed}}

\begin{document}

\title{Measurement-invisible quantum correlations in scrambling dynamics}

\author{Alan Sherry}
\email{alan.sherry@icts.res.in}
\affiliation{International Centre for Theoretical Sciences, Tata Institute of Fundamental Research, Bengaluru 560089, India}

\author{Sthitadhi Roy}
\email{sthitadhi.roy@icts.res.in}
\affiliation{International Centre for Theoretical Sciences, Tata Institute of Fundamental Research, Bengaluru 560089, India}

\begin{abstract}
Scrambling unitary dynamics in a quantum system transmutes local quantum information into a non-local web of correlations which manifests itself in a complex spatio-temporal pattern of entanglement. In such a context, we show there can exist three distinct dynamical phases characterised by qualitatively different forms of quantum correlations between two disjoint subsystems of the system. Transitions between these phases are driven by the relative sizes of the subsystems and the degree scrambling that the dynamics effects. Besides a phase which has no quantum correlations as manifested by vanishing entanglement between the parts and a phase which has non-trivial quantum correlations quantified by a finite entanglement monotone, we reveal a new phase transition within the entangled phase which separates phases wherein the quantum correlations are invisible or visible to measurements on one of the subsystems. This is encoded in the qualitatively different properties of the ensemble of states on one of the subsystems conditioned on the various measurement outcomes on the other subsystem. This provides a new characterisation of entanglement phases in terms of their response to measurements instead of the more ubiquitous measurement-induced entanglement transitions. Our results have implications for the kind of tasks that can be performed using measurement feedback within the framework of quantum interactive dynamics.
\end{abstract}

\maketitle

The spatio-temporal structure of the web of quantum correlations pervading a system, as encoded in the quantum entanglement between different parts of the system, is a cornerstone of modern quantum many-body physics.
It takes centre stage in a multitude of questions ranging from emergence of statistical mechanics in isolated quantum systems~\cite{nandkishore2015many,dalessio2016from,nahum2017quantum,garrison2018does}
to information scrambling in chaotic and holographic systems~\cite{hayden2007black,sekino2008fast,liu2014entanglement,liu2014entanglement1,hosur2016chaos,xu2019locality} and error correction in quantum computation~\cite{nielsen-chaung-book,kitaev2003fault,pastawski2015holographic,choi2020quantum,fan2021self,bravyi2024how}.

In the context of condensed matter and statistical physics, the emergence of entanglement as a central quantity of interest can be attributed to broadly two intertwined but complementary factors.
First, entanglement structure has emerged as a fundamentally novel paradigm for classifying phases of quantum matter, both in~\cite{calabrese2004entanglement,hamma2005ground,kitaev2006topological,levin2006detecting,swingle2010entanglement,eisert2010area,wen2017zoo} and out of equilibrium~\cite{calabrese2005evolution,calabrese2007entanglement,bauer2013area,nandkishore2015many,dalessio2016from,abanin2019colloquium,alet2018many,laflorencie2022entanglement,li2018quantum,skinner2019measurement,li2019measurement,Gullans2020Purification,choi2020quantum,fan2021self,nahum2021measurement,potter2022entanglement,fisher2023random}.
Second, the interface between condensed matter physics and quantum information science is increasingly blurring and it is entanglement that connects the two~\cite{eisert2006entanglement,laflorencie2016quantum}.
Questions pertinent to entanglement dynamics have taken on an ever more importance due to development of so-called noisy, intermediate scale quantum  devices which offer the possibility of a quantum interactive dynamics by implementing quantum circuits using the basic building blocks of unitary gates, quantum measurements, and feedback~\cite{preskill2018quantum,boixo2018characterising,smith2019simulating,ippoliti2021many,hoke2023measurement,fauseweh2024quantum}. In such settings, the entanglement structure provides proxy measures for quantifying the classical hardness of certain tasks and therefore distinguishes between tasks which offer a genuine possibility of quantum advantage from those which do not~\cite{schuch2008entropy,harrow2017nature,boixo2018characterising,arute2019quantum}.
It is therefore of immanent importance to understand and classify dynamical phases of quantum matter based on universal spatio-temporal patterns in their entanglement and how could the latter be manipulated to establish the entanglement as a resource for quantum computing.

The pattern of entanglement can be understood as a measure of how the information content of a state is distributed or {\it scrambled} non-locally across all the degrees of freedom of the system. 
This lies at the heart of entanglement growth going hand in hand with local thermalisation in scrambling dynamics; much of the information of a local subsystem is lost to or hidden amongst the rest of the system such that the state of former is well described by a much less structured thermal state. 
Dynamics which does not scramble information or does not effect thermalisation, such as in many-body localised systems or in measurement-unitary hybrid circuits with large enough measurement rate, lead to fundamentally different pattern of entanglement growth compared to scrambling dynamics. So, while it is clear that dynamical quantum phases can be classified based on their entanglement structure, one has to make specific what quantum correlations are being probed to shed light on the entanglement structure and what is the precise protocol being employed.

A minimal but natural protocol for probing the quantum correlations between two disjoint but extensive subsystems, $R$ and $S$ 
of an entire system $V\equiv R\cup S\cup E$ 
is to perform measurements on $S$ and ask how do the measurement outcomes feed back onto the state of $R$. 
Specifically, what are the properties of the ensemble 
\eq{
	{\cal E}_{RS}=\{\rho_R(o_S)\,;\,\, p(o_S)\}\,,
    \label{eq:ensemble}
}
where density matrix $\rho_R(o_S)$ describes the state of $R$ given the set of measurement outcomes $o_S$ and $p(o_S)$ is the corresponding probability.
Of course, this question is intricately interlaced with the genuinely quantum correlations between $R$ and $S$ as quantified via an entanglement monotone. 

\begin{figure*}
\includegraphics[width=\linewidth]{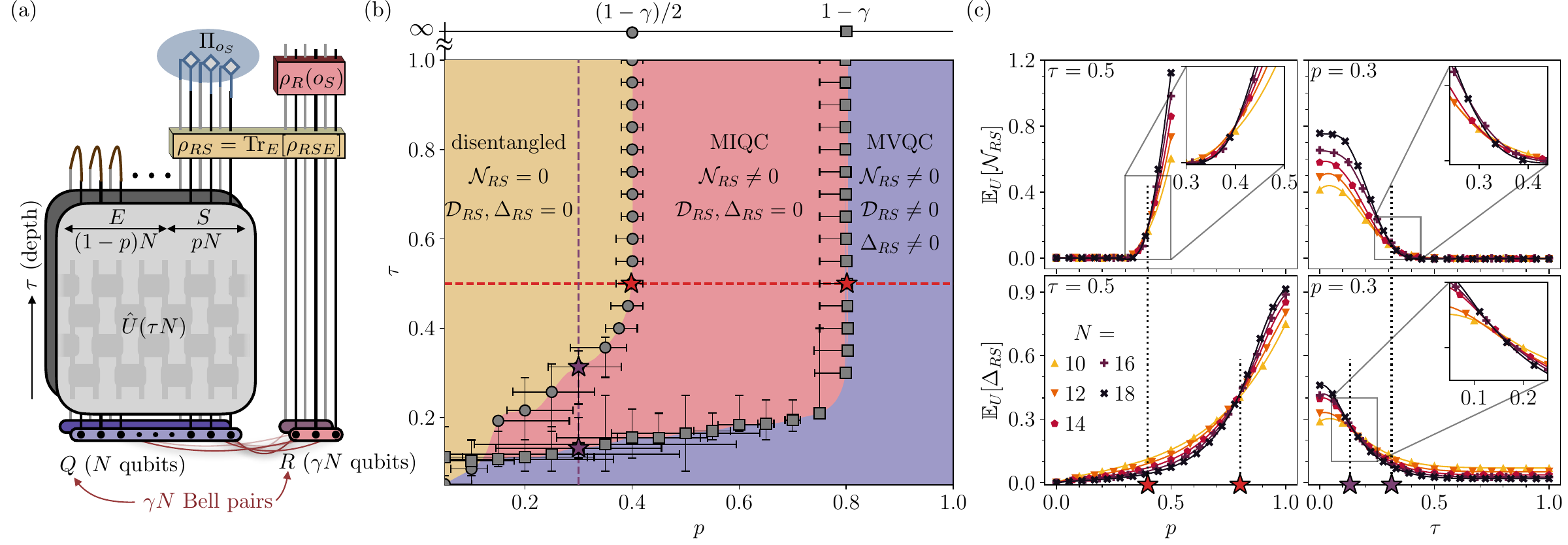}
\caption{
(a) Schematic of the setting. Initially, each of the $\gamma N$ qubits in $R$ forms a Bell pair with a randomly chosen qubit in $Q$ and the remaining $(1-\gamma)N$ qubits are in random product states. A unitary circuit of depth $\tau N$
 acts on $Q$. A subsystem $E$ of $Q$ is traced out, as indicated by the contractions, leading to state $\rho_{RS}$. Single-site, projective measurements are performed on the subsystem $S$ with outcome $o_S$ (the corresponding projection operator denoted by $\Pi_{o_S}$) to yield the state $\rho_R(o_S)$ of $R$. (b) The phase diagram in the $(p,\tau)$ plane for a fixed $\gamma=0.2$ \sr{obtained numerically except for the limiting case of $\tau\to\infty$ where the critical points are obtained analytically.} The yellow region corresponds to the phase where $R$ and $S$ are disentangled whereas the in red and purple regions they are entangled as measured via the logarithmic negativity ${\cal N}_{RS}$ in Eq.~\ref{eq:negativity}. The grey circles denote the entanglement transition points. The grey squares denote the measurement-invisibility transition within the entangled phase which separates the measurement-invisible (red) phase from the measurement-visible (purple) one which is diagnosed by ${\Delta}_{RS}$ defined in Eq.~\ref{eq:Delta-RS} or equivalently by ${\cal D}_{RS}$ in Eq.~\ref{eq:D-RS}. 
 The critical points in the limit of $\tau\to\infty$, at $p_{c,{\cal N}} = (1-\gamma)/2$ and $p_{c,{\cal D}}=1-\gamma$, in this case at 0.4 and 0.8 respectively are obtained analytically (see text for details).
 (c) Explicit data for the circuit averaged ${\cal N}_{RS}$ (top row) and $\Delta_{RS}$ (bottom row) for two representative slices (marked by the red horizontal and purple vertical dashed lines) of the phase diagram. The corresponding critical points are marked by the red and purple stars in both (b) and (c). Results in (b) and (c) correspond to the unitary being a Floquet Haar-random circuit with brickwork geometry. The details of how the critical points and error-bars are estimated is relegated to the Supp. Matt.~\cite{supp}.
 }
\label{fig:setting-pd}
\end{figure*}

In this work, we address this question for scrambling dynamics within the precise setting shown in Fig.~\ref{fig:setting-pd}(a). 
To establish if there is finite entanglement between $R$ and $S$, we use the logarithmic negativity~\cite{vidal2002computable,plenio2005logarithmic,eisert2006entanglement} denoted by ${\cal N}_{RS}$ as the state of $\rho_{RS}$ of $R\cup S$ will be mixed in general.
However, to understand if and how measurements on $S$ can probe the quantum correlations between $R$ and $S$ we study if the density matrices $\{\rho_R(o_S)\}$ in the ensemble \eqref{eq:ensemble} have a non-trivial distribution over $o_S$.

Based on these diagnostics we uncover a rich phase diagram (see Fig.~\ref{fig:setting-pd}(b)) in the parameter space spanned by the relative Hilbert-space dimensions ${D}_R$, ${D}_S$ and ${D}_E$, and the degree to which the dynamics effects scrambling within $S\cup E$.
There exists a critical surface which separates a disentangled phase with ${\cal N}_{RS} \to 0$ from an entangled one where ${\cal N}_{RS}\neq 0$ and is extensive~\footnote{Strictly speaking ${\cal N}_{RS} = 0$ does not imply absence of quantum entanglement; examples include bound entangled states~\cite{horodecki1998mixed,bruss2000construction}. \sr{However, in our setting, particularly at infinite time, there is no entanglement between $R$ and $S$ when ${\cal N}_{RS} = 0$ as it also corresponds to a decoupled phase.}}.
We also find that, in the former, the ensemble ${\cal E}_{RS}$ is concentrated strongly at the maximally mixed state, $\mathbb{I}_{{D}_R}$; it therefore follows trivially that the state $\rho_R(o_S)$ is independent of the measurement outcome $o_S$.
More interestingly, within the entangled phase, there exists another critical surface. 
This separates a phase where the ensemble ${\cal E}_{RS}$ continues to be concentrated around $\mathbb{I}_{{D}_R}$ despite $R$ and $S$ being entangled from a phase where the $\{\rho_R(o_S)\}$ has a non-trivial distribution over $o_S$.
In effect, in the former of the two entangled phases the state $\rho_R(o_S)$ is independent of the measurement outcome $o_S$ and hence we name this phase as a {\it measurement-invisible quantum correlated} (MIQC) phase. 
This entanglement phase diagram for scrambling dynamics, and in particular, the identification of the novel MIQC phase constitutes the central result of this work.

To put the aforementioned ideas and results on a concrete footing, we consider a setting (Fig.~\ref{fig:setting-pd}(a)) ubiquitously employed in quantum information theory~\cite{devetak2005private,horodecki2006quantum,abeyesinghe2009mother}.
It consists of a subsystem $R$ with $\gamma N$ qubits (with $0<\gamma<1$) and another subsystem $Q$ with $N$ qubits.
Initially, each qubit in $R$ forms a Bell pair with a randomly chosen qubit in $Q$, and the rest of the $(1-\gamma)N$ qubits in $Q$ are in random product states. 
This choice of initial condition is arguably the simplest which also ensures maximal entanglement between $R$ and $Q$.
A local Floquet unitary circuit, denoted by $\hat{U}(t)$, of depth $t = \tau N$ then acts on $Q$, with the resultant state denoted by $\rho_{RSE}$. 

While this unitary operation cannot change the total entanglement between $R$ and $Q$, it dynamically modifies the spatio-temporal structure of the entanglement as encoded in the quantum correlations between $R$ and any subsystem $S$ of $Q$ containing $pN$ qubits ($0<p<1$). 
We also denote by $E$ the remaining subsystem of $Q$ containing $(1-p)N$ qubits.
The parameter space of the model is therefore spanned by the three parameters $\gamma$, $p$, and $\tau$.

We quantify the entanglement between $R$ and $S$ via the logarithmic negativity~\cite{vidal2002computable,plenio2005logarithmic,eisert2006entanglement} defined as 
\eq{
{\cal N}_{RS} \equiv \ln\norm{\rho_{RS}^{\mathsf{T}_S}}_1\,,
\label{eq:negativity}
}
where $\rho_{RS} = \Tr_E[\rho_{RSE}]$ is the reduced density matrix of $R\cup S$, the superscript ${\mathsf{T}}_S$ denotes a partial transpose with respect to $S$, and $\norm{\cdot}_1$ denotes the trace norm. 
${\cal N}_{RS}$ acts as the `order parameter' for the phase transition separating the entangled phase from the disentangled.
A subtle point to note here is that the negativity is an upper bound to distillable entanglement; hence to show that in the disentangled phase $R$ and $S$ are indeed decoupled we also compute $\norm{\rho_{RS}-\rho_R\otimes\rho_S}_1$ where $\rho_{R(S)} = \Tr_{S(R)}\rho_{RS}$ and analytically show that it vanishes in the limit of $\tau\to\infty$.

In order to distinguish the two phases within the entangled phase, we consider the {\it projected} ensemble, ${\cal E}_{RS}$ defined in Eq.~\ref{eq:ensemble}. 
To construct the ensemble, projective measurements are performed on each qubit in $S$ in the computational basis such that the outcome $o_S$ is a string of $\pm 1$ of length $pN$.
Defining the projector onto the product state $\ket{o_S}$ as $\Pi_{o_S}$, ${\cal E}_{RS}$ in Eq.~\ref{eq:ensemble} is given by
\eq{
\rho_R(o_S) = \frac{\Tr_S[\Pi_{o_S}\rho_{RS}]}{p(o_S)}\,;\quad p(o_S) = \Tr[\Pi_{o_S}\rho_{RS}]\,.
\label{eq:rhoR(oS)-p(oS)}
}
In terms of this notation, the first moment of ${\cal E}_{RS}$ is the state of $R$, $\rho_R = \sum_{o_S}p(o_S)\rho_R(o_S) = \mathbb{I}_{{D}_R}/{D}_R$ where the second equality follows from the initial state being maximally entangled between $R$ and $Q$. 
Whether the elements of the set $\{\rho_R(o_S)\}$ have a non-trivial distribution over $o_S$ is most simply probed by
\eq{
{\cal D}_{RS} = \sum_{o_S}p(o_S) {\cal S}_{\rm rel}[\rho_R(o_S)|| \rho_R]\,,
\label{eq:D-RS}
}
where ${\cal S}_{\rm rel}(\rho||\phi)$ is the quantum relative entropy between the density matrices $\rho$ and $\phi$. 
${\cal D}_{RS}$ can be understood as a measure of fluctuation in the state of $R$ across the ensemble ${\cal E}_{RS}$; a vanishingly small value implies that $\{\rho_R(o_S)\}$ is strongly concentrated around $\rho_R$ and hence the measurement outcome $o_S$ is invisible to $\rho_R(o_S)$.
We ensure that the resulting measurement-invisibility transition is not an artefact of the choice of measure \eqref{eq:D-RS} by also studying another measure \eq{{\Delta}_{RS} = \sum_{o_S}p(o_S) \frac{\norm{\rho_R(o_S)-\rho_R}_1}{2}\,,\label{eq:Delta-RS}} which shows the same qualitative behaviour as ${\cal D}_{RS}$.

In the above, we have laid out the quantities, ${\cal N}_{RS}$ and ${\cal D}_{RS}$ which act as order parameters for the two transitions. We now turn towards explicit results for them using Floquet random unitary circuits (FRUC) \sr{with brickwork geometry} and a translation-invariant but quantum chaotic Floquet Ising chain. 
The results for the two models are qualitatively similar and we relegate those for the latter to the Supp. Matt.~\cite{supp}.
\sr{The unitary time-evolution operator for the FRUC is given by $\hat{U}(t)=U_F^t$ with the Floquet unitary $U_F = \prod_{i=1}^{L/2}u_{2i-1,2i}\prod_{i=1}^{L/2} u_{2i,2i+1}$ where $u_{i,i+1}$ is a $4\times 4$ Haar-random unitaries acting on qubits at sites $i$ and $i+1$ }


At $\tau=0$, the correlations between $R$ and $S$ are trivially measurement-visible for all sizes of $S$ as the state of a qubit in $R$ is completely slaved to the measurement outcome of its Bell pair partner qubit in $S$. This guarantees ${\cal D}_{RS}\neq 0$ for all $p$ and ${\cal N}_{RS}\neq 0$ is ensured by construction.
At low depth, $\tau\ll 1$, the scrambling of correlations has not been sufficiently effected by the unitary and their morphology remains similar to that of the initial state. Hence, ${\cal D}_{RS}$ and ${\cal N}_{RS}$ continue to be non-zero and the correlations between $R$ and $S$ continue to be measurement-visible for all sizes of $S$. 
At extensive depths, Fig.~\ref{fig:setting-pd}(b)-(c) shows that there exist two transitions as a function of $p$. A critical point $p_{c,\cal N}$ separates a phase where ${\cal N}_{RS}$ is extensive from one where ${\cal N}_{RS}\to 0$ as $N\to\infty$ indicated by the crossing of the data for different $N$. This can be understood as follows. At large depth the initial pattern of correlations between $R$ and $Q$ is heavily scrambled across $Q$. Tracing out a large subsystem $E$ of $Q$ therefore washes away the quantum correlations between $R$ and $S$ leading to vanishing entanglement between them for $p<p_{c,\cal N}$. This heuristic picture can be made rigorous using ideas of quantum decoupling as we do shortly.
Within the entangled phase there exists another critical point $p_{c,\cal D}(>p_{c,\cal N}$) which separates phases with ${\cal D}_{RS}$ extensive and vanishing respectively. For $p>p_{c,\cal D}$, the size of $S$ is large enough that sufficient information about the quantum correlations between $R$ and $S$ are retained in $\rho_{RS}$ for the phase to be measurement-visible as indicated by a non-vanishing ${\cal D}_{RS}$.
However, most interestingly, there exists the MIQC phase for $p_{c,\cal N}<p<p_{c,\cal D}$ where ${\cal D}_{RS}\to 0$ as $N\to\infty$ but ${\cal N}_{RS}$ is non-vanishing.
In this phase, $\mathcal{E}_{RS}$ is highly concentrated around the reduced density matrix $\rho_R$. As such, $\rho_R(o_S)$ does not depend non-trivially on the measurement outcome $o_S$ and the quantum correlations between $R$ and $S$ are therefore {\it measurement-invisible}.
This is despite there being non-vanishing entanglement between $R$ and $S$ and yet, one is unable to exploit the entanglement to engineer non-trivial ensembles for $R$ via back-action of measurements on $S$. This forms the essence of the MIQC phase.

Having established the three phases numerically along with a broad-brush view of their understanding, we now turn to a limit, namely of $t\to\infty$, where the two critical points can be obtained analytically. In this limit, $\hat{U}({t\to\infty})\to {\cal U}$ where ${\cal U}$ is a $2^N$-dimensional Haar-random unitary. We state the result at the outset that in this limit, the entanglement transition occurs at $p_{c,{\cal N}}=(1-\gamma)/2$ and the measurement-invisibility transition transition at $p_{c,{\cal D}}=1-\gamma$. In the following, we sketch how these critical points are obtained analytically and the details of the derivation are presented in the Supp. Matt.~\cite{supp}.

Using standard methods of Haar-averaging~\cite{weingarten1978asymptotic,collins2006integration}, denoted by $\mathbb{E}_{\cal U}$,
it can be shown that~\cite{choi2020quantum,supp}
\eq{
\mathbb{E}_\mathcal{U}[||\rho_{RS}-\tfrac{\mathbb{I}}{2^{(p+\gamma)N}}||_1]\leq
\Big[\tfrac{(2^{2pN}-1)(2^{(\gamma+1)N}-1)}{2^{2N}-1}\Big]^{1/2},
\label{eq:max-mix}
}
which implies that for $p<(1-\gamma)/2$, $\rho_{RS}$ is maximally mixed in the ${N\to\infty}$ limit and therefore decoupled, which then directly implies ${\cal D}_{RS},{\cal N}_{RS}=0$ in this parameter regime. 
\sr{An analytic calculation of ${\cal N}_{RS}$ is outside the scope of this paper. Although results for negativity exist for  random states \cite{PhysRevA.81.052312,shapourian2021negativity}, there is no {\it a priori} justification that the state $\ket{\psi_{RSE}}$ or the reduced state $\rho_{RS}$ is random.
We therefore use the mutual information between $R$ and $S$ as proxy for the entanglement between them. We will shortly justify why the mutual information and the negativity capture the entanglement transition equally well in this case.} The mutual information is defined as 
${\cal I}_{RS}\equiv\mathcal{S}_{{\rm vN},R}+\mathcal{S}_{{\rm vN},S}-\mathcal{S}_{{\rm vN},RS}$, with $\mathcal{S}_{{\rm vN},X}$ the von-Neumann entropy of $\rho_X$. 
While the details of the derivation can be found in the SM~\cite{supp}, for $p>(1-\gamma)/{2}$, the Haar-averaged mutual information can be lower bounded as
\eq{
\mathbb{E}_\mathcal{U}[{\cal I}_{RS}]&\geq\ln\left[\tfrac{2^{2(\gamma+p)N}-2^{2(\gamma+p-1)N}}{2^{2pN}+2^{(1+\gamma)N}-1-2^{(2p+\gamma-1)N}}\right]\,.
\label{eq:IRS-bound}
}
In the limit of $N\gg 1$ the RHS of Eq.~\ref{eq:IRS-bound} is approximately $(2p-1+\gamma)N$ and $2\gamma N$ for $p\lessgtr(1+\gamma)/2$ respectively. Crucially it is non-vanishing in the $N\to\infty$ limit 
which shows that $R$ and $S$ are correlated for $p>(1-\gamma)/2$.
Numerical results in Fig.~\ref{fig:setting-pd}(b)-(c) confirm that the logarithmic negativity shows a transition at $p_{c,\cal N}=(1-\gamma)/2$ for depth of $\tau\gtrapprox 0.5$. In fact, it is interesting to note that such depths are sufficient to observe the $t\to\infty$ limiting behaviour.
\sr{We can now justify why the mutual information is indeed a good proxy for the negativity. The disentangled phase, at $p<(1-\gamma)/2$ is, in fact, decoupled which naturally means not only are $R$ and $S$ disentangled, but also uncorrelated implying both ${\cal I}_{RS},{\cal N}_{RS}=0$. In the entangled phase however, the bound derived in Eq.~\ref{eq:IRS-bound} suggests that for $p>(1-\gamma)/2$, $R$ and $S$ become correlated. It therefore stands to reason that they also get entangled across $p_{c,{\cal N}}$, a result which is well supported by the numerical results as well.}
We therefore establish rigorously that (i) for $p<(1-\gamma)/2$, in the thermodynamic limit ${\cal D}_{RS},{\cal N}_{RS}=0$, and (ii) for $p>(1-\gamma)/2$, the mutual information ${\cal I}_{RS}$ is extensive. This shows that the critical point between the entangled and disentangled phases lies at $p_{c,\cal N}=(1-\gamma)/2$ for $\tau\gtrapprox 1$.

We next turn to the measurement-invisibility transition. In particular, we present analytic results for bounds on Haar-averaged ${\cal D}_{RS}$, defined in Eq.~\ref{eq:D-RS}. Recall that the initial state was maximally entangled between $R$ and $Q$, and the unitary $\hat{U}(t)$ acts only $Q$; consequently $\rho_R = \mathbb{I}/2^{\gamma N}$ for all $t$. This leads to 
\eq{
\mathcal{S}_{\rm rel}(\rho_R(o_s)||\rho_R)=\gamma N\ln 2-\mathcal{S}_{{\rm vN}}(\rho_R(o_s))\,.\label{eq:Srel-exp}
}
$\mathcal{S}_{\rm rel}(\rho_R(o_s)||\rho_R)$ can therefore be bounded by deriving bounds on $\mathcal{S}_{{\rm vN}}(\rho_R(o_s))$, which is in turn bounded by the R\'enyi entropies as
${\mathcal{S}_0(\rho)\geq \mathcal{S}_{\rm vN}(\rho)\geq \mathcal{S}_2(\rho)}$, where ${\cal S}_n(\rho)\equiv\frac{1}{1-n}\ln\rm{Tr}\rho^n$.

Defining $\tilde{\rho}_R(o_S)=\rho_R(o_s)/p(o_S)$ and upon Haar-averaging we have
\eq{
\mathbb{E}_{\cal U}[{\cal S}_2(\rho_R(o_S))] &= -\mathbb{E}_{\cal U}[\ln{\rm Tr}[\tilde{\rho}_R(o_S)]^2]+\mathbb{E}_{\cal U}[\ln p(o_S)^2]\nonumber\\
&\overset{N\gg 1}{\approx} -\ln\frac{\mathbb{E}_{\cal U} [{\rm Tr}[\tilde{\rho}_R(o_S)]^2]}{\mathbb{E}_{\cal U}[p(o_S)]^2}\equiv  \Lambda \,,
\label{eq:self-avg}
}
where the `quenched' average in the first line can be approximated by the `annealed' average in the second line due to the self-averaging property of Haar-random unitaries at large $N$~\cite{supp}. The inequality ${{\cal S}_{\rm vN}\ge {\cal S}_2}$ leads to 
\eq{
\mathbb{E}_{\cal U}[\mathcal{S}_{\rm rel}(\rho_R(o_s)||\rho_R)]\leq \gamma N\ln2 -\Lambda\,.
\label{eq:Srel-res}
}
A key point is after averaging, the result in Eq.~\ref{eq:self-avg}, denoted by $\Lambda$, is independent of $o_S$, such that the average over the measurement outcomes $o_S$ in Eq.~\ref{eq:Srel-res} can be performed trivially to yield
$\mathbb{E}_{\cal U}[{\cal D}_{RS}]\leq (\gamma N \ln 2 -\Lambda)$. As shown in the SM~\cite{supp}, $\Lambda$ can be analytically computed to yield
\eq{\mathbb{E}_{\cal U}[\mathcal{D}_{RS}]\leq \ln\left[\tfrac{2^{\gamma N}(2^{(1-\gamma) N}+2^{pN}-2^{(p-\gamma-1)N}-1)}{2^{(p-\gamma)N}+2^N-2^{-\gamma N}-2^{(p-1)N}}\right]\,.
\label{eq:EU-DRS}
}
For $p<1-\gamma$, the asymptotic value of the RHS of Eq.~\ref{eq:EU-DRS} decays exponentially in $N$, from which we conclude that $\mathcal{E}_{RS}$ is concentrated infinitely sharply around $\rho_R$ for $p<1-\gamma$ in the limit ${N\to\infty}$. This `proves' that the MIQC phase exists for $(1-\gamma)/2 < p < 1-\gamma$.
For $p>1-\gamma$, on the other hand, $\mathcal{E}_{RS}$ does not exhibit this concentration \textit{for any choice of unitary}. To prove this, we note,
\eq{
    \mathcal{S}_{{\rm vN}}(\rho_R(o_S))\leq \mathcal{S}_0(\rho_R(o_S))\leq  \min(1-p,\gamma)N\ln 2\,,
}
where the right inequality arises from the maximum Schmidt rank of the pure state $\ket{\psi_{RE}(o_S)}$ conditioned on the measurement outcome $o_S$. The choice of unitary is immaterial. The above inequality implies that, for $p>1-\gamma$
\begin{equation}
    \mathbb{E}_{\cal U}[\mathcal{D}_{RS}]\geq(\gamma+p-1)N\ln 2\,,
\end{equation}
which is extensive and hence forbids the existence of the MIQC phase for any $p>1-\gamma$ for any unitary. 
We have therefore proved that the measurement-invisibility transition occurs at $p_{c,\cal D}=1-\gamma$.
This completes the description of our analytical results for the two critical points and the dynamical phases in the $t\to\infty$ limit. 
These analytical results along with the numerical results (see Fig.~\ref{fig:setting-pd}(b)) discussed earlier allow us to chart out the entire phase diagram, shown in Fig.~\ref{fig:setting-pd}, with the MIQC phase being of central interest.
A few remarks about implications of the MIQC phase are therefore in order.

A fallout of the MIQC phase is that joint distributions of measurement outcomes of two observables, one supported exclusively on $R$ and the other on $S$, with the latter a tensor product of single site operators over $S$, factorise despite $R$ and $S$ being entangled. 
To see this, consider an observable $A_{R(S)}$ supported exclusive on $R(S)$. 
The joint probability that the measurements yield eigenvalues $a_R$ and $a_S$ is given by
\eq{
P(a_R,a_S) = {\rm Tr}[(\Pi^{A_R}_{a_R}\otimes\Pi^{A_S}_{a_S})\rho_{RS}]\,,
}
where $\Pi_{a_{R(S)}}^{A_{R(S)}}$ is the projector onto the eigenstate of $A_{R(S)}$ with eigenvalue $a_{R(S)}$. 
The concentration of ${\cal E}_{RS}$ around $\rho_R$ in the MIQC phase implies that the state of $R$ given a measurement outcome $a_S$, $\rho_R(a_S)$, is the same as $\rho_R$ for $N\to\infty$. 
One therefore has 
\eq{
P(a_R|a_S)={\rm Tr}[\rho_R(a_S)\Pi_{a_R}^{A_R}]={\rm Tr}[\rho_R\Pi_{a_R}^{A_R}] = P(a_R)\,,
}
from which it immediately follows that 
\eq{
P(a_R,a_S)=P(a_R|a_S)P(a_S) = P(a_R)P(a_S)\,,
}
and hence the aforementioned factorisation.

The existence of the MIQC phase also has an interesting consequence in the context of information theory akin to steering ~\cite{Schrodinger_1935,Schrodinger_1936,wiseman2007steering}. Consider a setting where Alice $(A)$ and Bob $(B)$ share a bipartite state which is entangled. Each of them can perform measurements on their respective parts and communicate classically and $B$ can perform tomography on his part of the state. $A$'s goal is to convince $B$ that the shared state is correlated. If the state shared between $A$ and $B$ is in the MIQC phase, $B$'s tomography yields a state which identical for every measurement outcome reported by $A$. The state then appears uncorrelated to $B$ and $A$ fails in their task. Note that this is a strictly stronger condition than \textit{unsteerability} as $A$'s failure to convince $B$ that the state is correlated would imply a failure to convince that the state is entangled. While bipartite, entangled pure states are necessarily steerable~\cite{gisin1991bell,gisin1992maximal,popescu1992generic}, the MIQC phase can therefore be viewed as an explicit example of unsteerable but entangled mixed states~\cite{wiseman2007steering}.

To summarise, we have discovered a new dynamical phase and a phase transition in scrambling dynamics in terms of the entanglement structure between two disjoint subsystems. The distinct entanglement structures manifest themselves in appropriate entanglement measures as well as in how one of the subsystems responds to projective measurements on the other. The centrepiece result is the MIQC phase where the two subsystems are entangled and yet, measurements on one of them are invisible to the other.

The discovery of the novel MIQC phase also raises several questions. 
One such question, of fundamental nature, is if, and to what extent, does the measurement-invisibility transition probe or distinguish genuine multipartite volume-law entangled states from those where the volume-law entanglement arises from there being few-partite entanglement between an extensive number finite subsystems. 
A closely related question is can the measures discussed in this work characterise the spatial structure of quantum information in a state.
Ideal playgrounds to answer these questions are non-scrambling or anomalously scrambling dynamics such as measurement-unitary hybrid circuits~\cite{li2018quantum,skinner2019measurement,li2019measurement,Gullans2020Purification,vijay2020measurement,choi2020quantum,fan2021self,bao2024finite} or many-body localised dynamics~\cite{bardarson2012unbounded,serbyn2013universal,chen2017out,deng2017logarithmic,maccormack2021operator} where the structure of information spreading is known to be qualitatively different from thermalising scrambling dynamics~\cite{nahum2017quantum,nahum2018operator,rakovszky2018diffusive,khemani2018operator}.

Understanding the implications of the MIQC phase and the entanglement structure therein on measurement-based protocols for quantum state preparation and stabilising phases with exotic quantum order~\cite{roy2019measurement,noel2022measurement,wampler2022stirring,mcginley2022absolutely,herasymenko2023measurement,friedman2023measurement}, particularly with long-ranged entanglement~\cite{tantivasadakarn2024long,iqbal2024topological} is another question of immediate future interest. \sr{In particular, the presence of an MIQC phase, say due to the presence of an environment, can lead to dramatic failure of protocols that are based on measurement operations on ancillary degrees of freedom, wherein the feedback from the ancillaries on the system is used to manipulate the state of the system. If the state of the system is insensitive to the measurements on the ancillaries, the entanglement between the system and ancillaries is not sufficient, and the protocol fails.}

\begin{acknowledgments}
We acknowledge support from SERB-DST, Government of India under Grant No. SRG/2023/000858, from the Department of Atomic Energy, Government of India, under Project No. RTI4001, and from a Max Planck Partner Group grant between ICTS-TIFR, Bengaluru and MPIPKS, Dresden.
\end{acknowledgments}

\bibliography{refs}

\clearpage

\setcounter{equation}{0}
\setcounter{figure}{0}
\setcounter{page}{1}
\renewcommand{\theequation}{S\arabic{equation}}
\renewcommand{\thefigure}{S\arabic{figure}}
\renewcommand{\thesection}{S\arabic{section}}
\renewcommand{\thepage}{S\arabic{page}}

\onecolumngrid
\begin{center}
{\large{\bf Supplementary Material: Measurement-invisible quantum correlations in scrambling dynamics}}\\
\medskip

Alan Sherry and Sthitadhi Roy\\
{\small{\it International Centre for Theoretical Sciences, Tata Institute of Fundamental Research, Bengaluru 560089, India}}
\end{center}
\bigskip

The supplementary material consists of three sections. In Sec.~I, we present numerical results for the Floquet mixed-field Ising chain (FMFIC). In Sec.~II, we provide some details pertaining to the numerical computation of the phase diagrams for the FRUC and FMFIC, following which Sec.~III is devoted to the details of the analytical computation of the critical points in the limit of $t\to\infty$.

\section{I. Numerical results for the Floquet mixed-field Ising chain}
\begin{figure*}[htb]
    \includegraphics[width=0.48\linewidth]{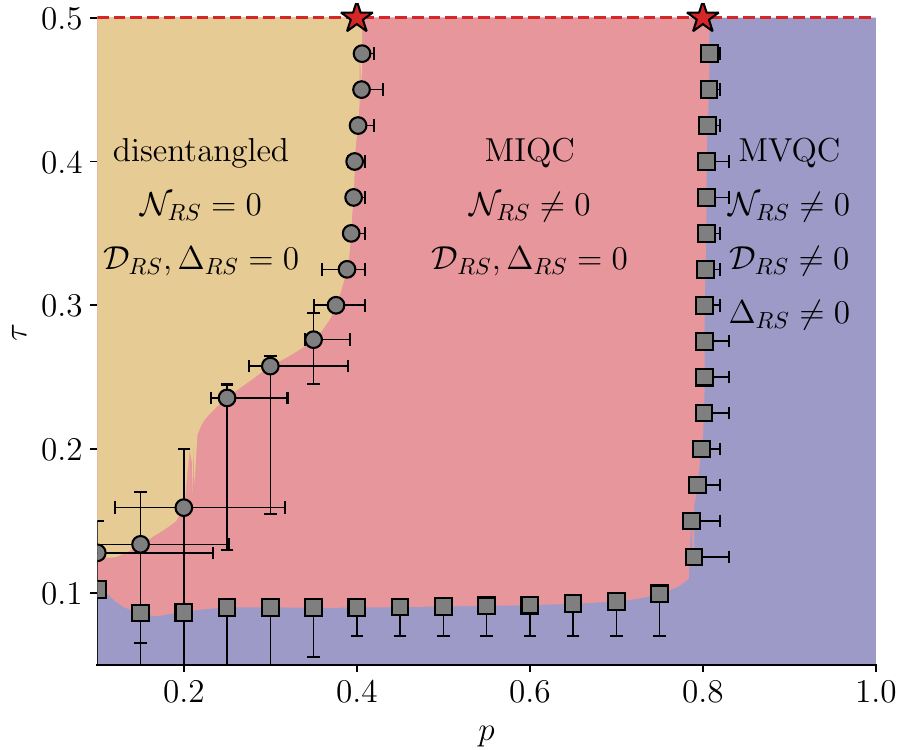}
     \includegraphics[height=0.4\linewidth,width=0.3\linewidth]{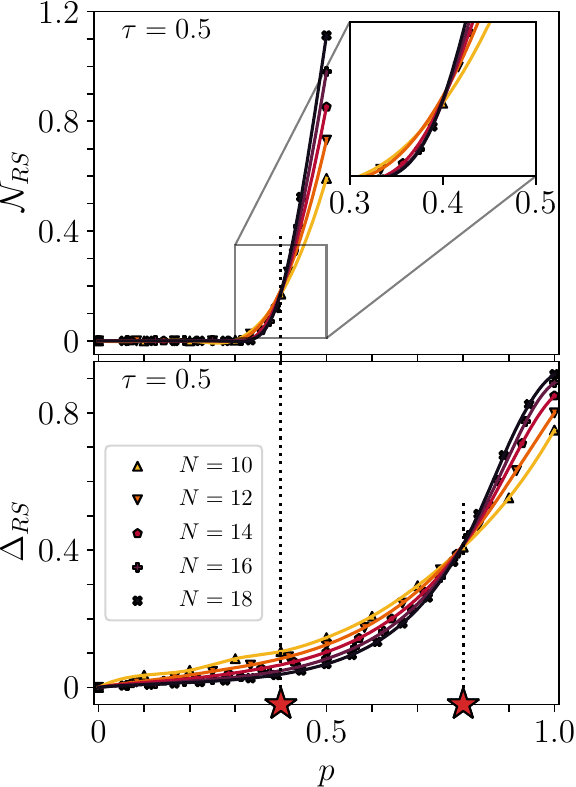}
    \caption{\textbf{Left}: The phase diagram for the Floquet mixed field Ising chain \eqref{eq:FMFIM} for $\gamma=0.2$.The critical points in the limit of $\tau\to\infty$ are $p_{c,\mathcal{N}}=0.4$ and $p_{c,\mathcal{D}}=0.8$. Note the similarity of the phase diagram and in particular the quantitative similarity in the limit $\tau\sim O(1)$ with that of the FRUC in Fig.~\ref{fig:setting-pd}(b). 
    \textbf{Right}: Explicit data for $\mathcal{N}_{RS}$ (top) and $\Delta_{RS}$ (bottom) for depth $\tau=1$. The corresponding critical points, marked by the red stars, are also marked similarly in the phase diagram (left).}
    \label{fig:FMFIM-phase}
\end{figure*}

In this section, we present numerical results for ${\cal N}_{RS}$ and $\Delta_{RS}$ for the FMFIC. 
These results, shown in Fig.~\ref{fig:FMFIM-phase}, complement those for the FRUC (presented in the main text in Fig.~\ref{fig:setting-pd}). As is customary with Floquet systems, we consider stroboscopic times $t$ with the time-evolution operator $U(t) = U_F^t$ where $U_F$ is the time-evolution operator over one period. The latter for the specific FMFIC we consider is given by
\eq{
U_{F} = e^{-iH_+}e^{-iH_-}\,\quad{\rm with}\quad H_{\pm} = \sum_{i=1}^N[h_X(1\pm\delta)X_i + h_YY_i + J X_iX_{i+1}]\,,
\label{eq:FMFIM}
}
where $X_i$ and $Y_i$ denote the Pauli-$x$ and Pauli-$y$ operators for the spin-1/2 at site $i$. We take the parameters to be $(h_X,h_Y,J,\delta)=(0.809,0.9045,1,0.5)$ where the model is known to be quantum chaotic.
For consistency of notation we define $t=\tau N$ as the depth of the time-evolution unitary.
The phase diagram obtained for this model (Fig.~\ref{fig:FMFIM-phase}) is qualitatively similar to that of the FRUC and in fact, in the limit of $t\gg N$ the critical points approach the values of $p_{c,{\cal N}}=(1-\gamma)/2$ and $p_{c,{\cal D}}=(1-\gamma)$ as obtained numerically for the FRUC in the limit of $t\gtrapprox N$ and analytically for Haar-random unitaries.
The key takeaway from these results is that the presence of the entanglement transition and the measurement-invisibility transition is a robust feature of scrambling dynamics effected in quantum chaotic systems. In particular, note that the FMFIC, \eqref{eq:FMFIM}, is translation invariant and hence it presents compelling evidence that the transitions are not features of the averaging over Haar-random circuits.

\section{II. Details about obtaining the phase diagrams and estimation of error-bars}
In this section, we list out some of the details of, and procedures employed, in the numerical computation for the sake of completeness.
\begin{itemize}
\item The total system size whose state needs to be kept track of is $N_{\rm {tot}}=(1+\gamma)N$ qubits. However, note that for most choices of $\gamma$ and $N$ within accessible system sizes $\gamma N$ is not an integer. We circumvent this problem simulating systems with $N_{>} = N + \lceil \gamma N\rceil$ and $N_< = N+\lfloor \gamma N\rfloor$ qubits and averaging all results with weights $\eta_>$ and $\eta_< (=1-\eta_>)$ such that $N_{\rm {tot}} = \eta_>N_>+\eta_<N_<$.

\item Given that we have discrete $N$ we can only obtain results for a discrete set of $N+1$ values of $p = 0, 1/N, 2/N,\cdots,1$. Similarly, since we use a FRUC with a brickwork geometry and a Floquet Ising chain, the dynamics is naturally in discrete (stroboscopic) times and similarly we can  track only discrete $\tau = 0, 1/N, 2/N,\cdots, t/N$ where we $t$ is the number of steps for which we run the dynamics. In such a situation we estimate the critical points as follows. For a given $N$, we consider ${\cal N}_{RS}$ and $\Delta_{RS}$ as discrete two-dimensional functions on a $(N+1)\times (t+1)$ grid and generate an interpolating function of each of the two quantities, $f_{{\cal N},N}(p,\tau)$ and $f_{{\Delta},N}(p,\tau)$ respectively. We then consider the functions $f_{{\cal N},N}(p,\tau)$ and $f_{{\Delta},N}(p,\tau)$ for several $N$ at an arbitrary $p$ and $\tau$. If $f_{{\cal N},N}(p,\tau)$ grows with $N$ on an average we consider that $(p,\tau)$ to be in the entangled phase and if it decays or stays constant with $N$ (within numerical precision) we consider that phase to disentangled. This locus of points separating these two behaviours gives us the critical points for the entanglement transition. An identical procedure with $f_{{\Delta},N}(p,\tau)$ yields the critical points for the measurement-invisibility transition.

\item The procedure we employ for estimating the error-bars in the phase diagram is as follows. For a given $\tau$, the critical point $p_{c,{\cal N}}(\tau)$ is estimated, as discussed above, as the point which separates a average growing trend of $f_{{{\cal N}},N}(p,\tau)$ with $N$ from an average decaying or constant trend with $N$. The horizontal error-bars (errors in $p_{c,{\cal N}}(\tau)$) are then estimated by locating the $p$ where $f_{{{\cal N}},N}(p,\tau)$ necessarily grows and necessarily decays with $N$ monotonically with former yielding the upper error and the latter the lower error. Repeating the same procedure for a fixed $p$ yields the vertical error-bars denoting the errors in $\tau_{c,{\cal N}}(p)$. The error-bars in $p_{c,\Delta}(\tau)$ and $\tau_{c,\Delta}(p)$ are estimated likewise by considering the behaviour of $f_{{\Delta},N}(p,\tau)$ with $N$.
\end{itemize}


\section{III. Derivation and details of results in the $t\to\infty$ limit}

\subsection{III A. Derivation of inequality in Eq.~\ref{eq:IRS-bound} in main text}

In this section, we provide the details of the bounds on the mutual information ${\cal I}_{RS}$ which is given by
\eq{
{\cal I}_{RS} = \mathcal{S}_{{\rm vN},R}+\mathcal{S}_{{\rm vN},S}-\mathcal{S}_{{\rm vN},RS}\,.
\label{eq:IRS-supp}
}
The starting point to derive the bound in Eq.~\ref{eq:IRS-bound} is to note that the $n^\text{th}$ R{\'e}nyi entropies, defined as ${\cal S}_{n,X}=\frac{1}{1-n}\ln\Tr[\rho_X^n]$ are a monotonic function of $n$ with ${\cal S}_{n+1,X}\leq {\cal S}_{n,X}$. Identifying that the $\lim_{n\to 1}{\cal S}_{n,X} = {\cal S}_{{\rm vN},X}$we have the bounds
%
%
\begin{equation}
         {\cal S}_{2,X}\leq {\cal S}_{{\rm vN},X}\leq {\cal S}_{0,X}
         \label{eq:Ap1}
     \end{equation} 
For a Haar-random unitary ${\cal U}$, we can  compute exactly the \textit{annealed} averages of the second R{\'e}nyi entropy ${{\cal S}^\text{ann}_{2,X}\equiv-\ln\mathbb{E}_\mathcal{U} [\Tr\rho^2_X]}$ using the identity
\begin{equation}
\mathbb{E}_\mathcal{{\cal U}}[{\cal U}^{\otimes 2}\otimes {\cal U}^{\dagger\otimes 2}]=\sum_{\sigma,\tau\in S_2}\chi_\sigma^{\otimes N}\otimes\chi_\tau^{\otimes N}\text{Wg}(\sigma\tau^{-1},2^N)
\label{eq:3}
\end{equation}
where $\sigma$ and $\tau$ are elements of the permutation group of order $2$, $\chi_{\sigma}=\sum_{i_1,i_2}\ket{i_{\sigma(1)}i_{\sigma(2)}}\bra{i_1i_2}$ is its representation in the replicated single-site Hilbert space $\{\ket{i}\}_{i=0,1}$, and Wg is the Weingarten function. Since $-\ln$ is a convex function, Jensen's inequality yields
 \begin{equation}
 \begin{split}
      -\ln\mathbb{E}_\mathcal{U}[\Tr[\rho_X^2]]&\leq \mathbb{E}_\mathcal{U}[-\ln\Tr[\rho_X^2]]\\
      \implies {\cal S}_{2,X}^\text{ann}&\leq \bar{{\cal S}}_{2,X}
 \end{split}
 \label{eq:Ap2}
 \end{equation}
If the state of $X\cup\bar{X}$ is pure,
its Schmidt rank is bounded above by $2^{\min(|X|,|\bar{X}|)}$, where $|X|$ denotes the number of qubits in subsystem $X$. Since ${\cal S}_{0,X}$ is the log of the Schmidt rank, we have the bound $\bar{S}_{0,X}\leq \min(|X|,|\bar{X}|)\ln 2$.  This, in addition to the inequalities in Eq.~\ref{eq:Ap1} and Eq.~\ref{eq:Ap2} gives us the combined inequality
\begin{equation}
     {\cal S}_{2,X}^\text{ann}\leq \bar{{\cal S}}_{2,X}\leq \bar{{\cal S}}_{{\rm vN},X}\leq \bar{{\cal S}}_{0,X}\leq \min(|X|,|\bar{X}|)\ln 2
     \label{eq:A1}
\end{equation}
The inequalities in  Eq.~\ref{eq:A1} imply that ${\cal S}_{{\rm vN},S}\ge S^{\rm ann}_{2,S}$ and $-{\cal S}_{{\rm vN},RS}\ge -\min(\gamma+p,1-p)N\ln 2$ for the first and third terms in the RHS of Eq.~\ref{eq:IRS-supp} respectively whereas ${\cal S}_{\rm vN,R}=\gamma N\ln 2$ is due to the fact that the unitary acts only on $Q (=S\cup E)$. Putting these together, we obtain a lower bound on the average mutual information between $R$ and $S$
\begin{equation}
    \mathbb{E}_\mathcal{U}[{\cal I}_{RS}]\geq {\cal S}^\text{ann}_{2,S}-\min(\gamma+p,1-p)N\ln 2+\gamma N\ln 2\,,
    \label{eq:A2}
\end{equation}
and we are then left with computing ${\cal S}^\text{ann}_{2,S}$.
To do so, we note that the purity of $\rho_S$ can be written as 
\begin{equation}
    \Tr\rho_S^2=\Tr[\mathsf{SWAP}_S(\mathbb{I}_R\otimes \mathcal{U})^{\otimes 2}\rho_0^{\otimes 2}(\mathbb{I}_R\otimes \mathcal{U}^\dagger)^{\otimes 2}]
    \label{eq:Ap3}
\end{equation}   
where $\mathbb{I}_R$ is the identity operator in the Hilbert space of $R$, ${\cal U}$ is the Haar-random unitary acting on $Q$, and  $\mathsf{SWAP}_S$ is the SWAP operator acting on the doubled Hilbert space of $S$. The latter can be formally written as 
\eq{
\mathsf{SWAP}_S=\otimes_x\chi_{g_x}\,,\quad{\rm with}\quad
    g_x=\begin{cases}
    (12)=g,&x\in S\\
    \text{identity}=e,&x\in\bar{S}
    \end{cases}\,,
}
where $g_x\in S_2$ labels the permutation on site $x$. $\rho_0$ is initial state, which is a tensor product of $\gamma N$ Bell-pair states $\frac{\ket{00}+\ket{11}}{\sqrt{2}}$ and $\ket{0}^{\otimes(1-\gamma)N}$ states, where $\ket{0}$ and $\ket{1}$ are eigenstates of the Pauli-$z$ operator.    
From Eq.~\ref{eq:3} and Eq.~\ref{eq:Ap3} we have,
\begin{equation*}
\begin{split}
\mathbb{E}_{\cal U}[\Tr[\mathsf{SWAP}_S(\mathbb{I}_R\otimes \mathcal{U})^{\otimes 2}\rho_0^{\otimes 2}(\mathbb{I}_R\otimes \mathcal{U}^\dagger)^{\otimes 2}]]
    =  \sum_{\sigma,\tau\in S_2}\text{Wg}(\sigma\tau^{-1},2^N)2^{-2\gamma N }\prod_{x\in R}\Tr\{\chi_{g_x}\chi_\tau\}\prod_{x\in SE}\Tr\{\chi_{g_x}\chi_\sigma\}\,.
\end{split}
\end{equation*}
Noting that $\Tr[\chi_\sigma\chi_\tau]=2^{|\sigma^{-1}\tau|}$ and $\text{Wg}(\sigma\tau^{-1},2^N)= \frac{\delta_{\sigma,\tau}-2^{-N}\delta_{g\sigma,\tau}}{2^{2N}-1}$, we have
\begin{equation}
\begin{split}
    \mathbb{E}_\mathcal{U}[\Tr[\rho_{S}^2]]
    &= \frac{2^{-2\gamma N}}{2^{2N}-1}\sum_{\sigma,\tau\in S_2}(\delta_{\sigma,\tau}-2^{-N}\delta_{g\sigma,\tau})2^{\gamma N|\tau|}2^{pN|g\sigma|}2^{(1-p)N|\sigma|}\\
     &= \frac{2^{-2\gamma N}}{2^{2N}-1}(2^{(\gamma+1+p)N}+2^{(2\gamma+2-p) N}-2^{(\gamma+1-p)N}-2^{(2\gamma+p)N})\,.
\end{split}
\label{eq:A4}
\end{equation} 
Using Eq.~\ref{eq:A4} in Eq.~\ref{eq:A2}, we have the inequality
\begin{equation}
    \mathbb{E}_\mathcal{U}[{\cal I}_{RS}]\geq -\ln \Bigg[ \frac{2^{N(1-3\gamma-p)}}{2^{2N}-1}(2^{(\gamma+1+p)N}+2^{(2\gamma+2-p) N}-2^{(\gamma+1-p)N}-2^{(2\gamma+p)N})
    \Bigg]
    \label{eq:A7}
\end{equation}
which is the inequality \eqref{eq:IRS-bound} in the main text.

\subsection{III B. Derivation of inequality in Eq.~\ref{eq:max-mix} in main text}

We present here some details of the derivation of the inequality in Eq.~\ref{eq:max-mix} involving the trace distance between $\rho_{RS}$ and the maximally mixed state $\mathbb{I}/2^{(p+\gamma)N}$ (also see \cite{choi2020quantum}). Consider the operator $(\rho_{RS}-\mathbb{I}/2^{(p+\gamma)N})$ and note that it possesses a real eigenspectrum $\{\lambda_i\}_{i=1..2^{(p+\gamma)N}}$. Using the Cauchy-Schwarz inequality,
\begin{equation}
 \begin{split}
\Big(\sum_i|\lambda_i|\Big)^2&\leq \sum_i \lambda_i^2\times2^{(p+\gamma)N}\\
\implies||\rho_{RS}-\mathbb{I}/2^{(p+\gamma)N}||^2_1&\leq2^{(p+\gamma)N}\Tr[(\rho_{RS}-\mathbb{I}/2^{(p+\gamma)N})^2]\\
&=2^{(p+\gamma)N}\Big(\Tr[\rho^2_{RS}]-2^{-(p+\gamma)N}\Big)
 \end{split}   
\end{equation}
To place bounds on the unitary-averaged trace distance, we note
\begin{equation}
\begin{split}
    \mathbb{E}_\mathcal{U}\Big[||\rho_{RS}-\mathbb{I}/2^{(p+\gamma)N}||_1\Big]&\leq\sqrt{\mathbb{E}_\mathcal{U}\Big[||\rho_{RS}-\mathbb{I}/2^{(p+\gamma)N}||_1^2\Big]}\\
    &\leq\sqrt{2^{(p+\gamma)N}\mathbb{E}_\mathcal{U}\Big[\Tr[\rho^2_{RS}]\Big]-1}\,,
\end{split}
\label{eq:td-supp}
\end{equation}
and we are left with computing $\mathbb{E}_{\cal U}[{\rm Tr}[\rho_{RS}^2]]$.
The computation is exactly analogous to the computation of $\mathbb{E}_{\cal U}[{\rm Tr}[\rho_{S}^2]]$ in the previous section and yields,
\begin{equation}
    \begin{split}
    \mathbb{E}_\mathcal{U} [\Tr[\rho_{RS}^2]]&=\frac{2^{-2\gamma N}}{2^{2N}-1}\sum_{\sigma,\tau\in S_2}(\delta_{\sigma,\tau}-2^{-N}\delta_{g\sigma,\tau})2^{\gamma N|g\tau|}2^{pN|g\sigma|}2^{(1-p)N|\sigma|}\\
    &=\frac{2^{-2\gamma N}}{2^{2N}-1}(2^{(2\gamma+p+1)N}+2^{(\gamma+2-p)N}-2^{(\gamma+p)N}-2^{(2\gamma-p+1)N})\,.
\end{split}
\label{eq:A5}
\end{equation}
Using Eq.~\ref{eq:A5} in Eq.~\ref{eq:td-supp} we obtain the inequality
\begin{equation}
\mathbb{E}_\mathcal{U}\Big[||\rho_{RS}-\mathbb{I}/2^{(p+\gamma)N}||_1\Big]\leq\sqrt{\frac{2^{(\gamma+2p-1)N}+2^{-2N}-2^{(2p-2)N}-2^{(\gamma-1)N}}{1-2^{-2N}}}\,,  
\end{equation}
which is nothing but the result in  \eqref{eq:max-mix} of the main text.

\subsection{III C. Derivation of inequality in Eq.~\ref{eq:EU-DRS} in the main text}

In Eq.~\ref{eq:EU-DRS} in the main text, the key quantity in the result was the $\Lambda$ defined as $\Lambda\equiv-\ln\frac{\mathbb{E}_{\cal U} [{\rm Tr}[\tilde{\rho}_R(o_S)]^2]}{\mathbb{E}_{\cal U}[p(o_S)]^2}$. In this section, we show explicitly its computation.
We start by noting that 
\begin{equation}
\mathbb{E}_\mathcal{U}[p(o_S)^2]=\mathbb{E}_\mathcal{U}[\Tr\{(I_R\otimes\ket{o_{S}}\bra{o_{S}}\otimes I_E)^{\otimes 2}(I_R\otimes \mathcal{U})^{\otimes 2}\rho_0^{\otimes 2}(I_R\otimes \mathcal{U}^\dagger)^{\otimes 2}\}]
    \label{eq:Bp1}
\end{equation}
The post-measurement state of $S$ $\ket{o_S}=\otimes_{x\in S} \ket{o_x}$, where $o_x\in\{0,1\}$ enumerates over the single-site basis states.
Using Eq.~\ref{eq:3} in Eq.~\ref{eq:Bp1} yields
\begin{equation}
    \begin{split}
        \mathbb{E}_\mathcal{U}[p(o_S)^2]&= \sum_{\sigma,\tau\in S_2}\text{Wg}(\sigma\tau^{-1},2^N)2^{-2\gamma N }\prod_{x\in R}\Tr\{\chi_{e}\chi_\tau\}\prod_{x\in E}\Tr\{\chi_{e}\chi_\sigma\}\prod_{x\in S}\Tr(\{\ket{o_x}\bra{o_x})^{\otimes 2}\chi_\sigma\}\\
        &=\frac{2^{-2\gamma N}}{2^{2N}-1}\sum_{\sigma,\tau\in S_2}(\delta_{\sigma,\tau}-2^{-N}\delta_{g\sigma,\tau})2^{\gamma N|\tau|}2^{(1-p)N|\sigma|}\\
     &= \frac{2^{-2\gamma N}}{2^{2N}-1}(2^{(\gamma+1-p)N}+2^{(2\gamma+2-2p) N}-2^{(\gamma+1-2p)N}-2^{(2\gamma-p)N})\,.
     \end{split}
     \label{eq:EU-possq-supp}
     \end{equation}
     Similarly, we obtain
     \begin{equation}
         \begin{split}        \mathbb{E}_\mathcal{U}[\Tr[\tilde{\rho}_R(o_S)]^2]&=\mathbb{E}_\mathcal{U}\Tr\{\mathcal{S}_{2,R}(I_{RE}\otimes\ket{o_{S}}\bra{o_{S}})^{\otimes 2}(I_R\otimes \mathcal{U})^{\otimes 2}\rho_0^{\otimes 2}(I_R\otimes \mathcal{U}^\dagger)^{\otimes 2}(I_{RE}\otimes \ket{i_{S}}\bra{i_{S}})^{\otimes 2}\}\\
&=\sum_{\sigma,\tau\in S_2}\text{Wg}(\sigma\tau^{-1},2^N)2^{-2\gamma N }\prod_{x\in R}\Tr\{\chi_{g}\chi_\tau\}\prod_{x\in E}\Tr\{\chi_{e}\chi_\sigma\}\prod_{x\in S}\Tr(\{\ket{o_x}\bra{o_x})^{\otimes 2}\chi_\sigma\}\\
&=\frac{2^{-2\gamma N}}{2^{2N}-1}\sum_{\sigma,\tau\in S_2}(\delta_{\sigma,\tau}-2^{-N}\delta_{g\sigma,\tau})2^{\gamma N|g\tau|}2^{(1-p)N|\sigma|}\\
     &= \frac{2^{-2\gamma N}}{2^{2N}-1}(2^{(\gamma+2-2p)N}+2^{(2\gamma+1-p) N}-2^{(\gamma-p)N}-2^{(2\gamma+1-2p)N})\,.
         \end{split}
         \label{eq:EU-tr-rhosq-supp}
     \end{equation}
     
Using Eq.~\ref{eq:EU-possq-supp} and Eq.~\ref{eq:EU-tr-rhosq-supp} in the definition of $\Lambda$ we obtain
\begin{equation}
   \Lambda=-\ln\Bigg[\frac{2^{(\gamma+2-2p)N}+2^{(2\gamma+1-p) N}-2^{(\gamma-p)N}-2^{(2\gamma+1-2p)N}}{2^{(\gamma+1-p)N}+2^{(2\gamma+2-2p) N}-2^{(\gamma+1-2p)N}-2^{(2\gamma-p)N}}\Bigg]\,,
    \label{eq:A14}
\end{equation}
which directly leads to the inequality in Eq.~\ref{eq:EU-DRS}.

\subsection{III D. Self-averaging for Haar random unitaries: justification for Eq.~\ref{eq:self-avg} in the main text}
\begin{figure*}[htb]
    \includegraphics[width=0.8\linewidth]{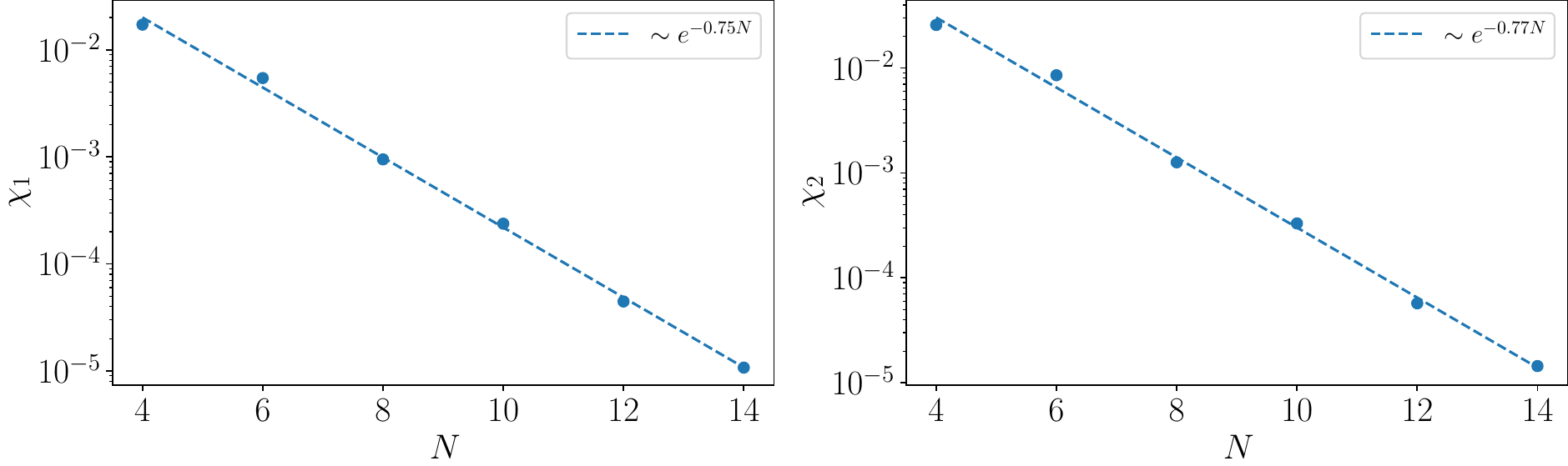}
    \caption{$\chi_1$ (left) and $\chi_2$ (right) for for $2^N$ dimensional Haar random unitaries. The (exponential) decay of the functions with $N$ provide credible justification for Eq.~\ref{eq:self-avg} in the main text.}
    \label{fig:self-avg}
\end{figure*}

In Eq.~\ref{eq:self-avg} of the main text, we replaced the quenched average over the Haar-random unitaries by the annealed average in the limit of $N\gg1$ for any measurement readout $o_S$ by appealing to the self-averaging property of large Haar-random unitary matrices,
\begin{equation}
    \begin{split}
       \mathbb{E}_{\cal U}[\ln{\rm Tr}[\tilde{\rho}_R(o_S)]^2]&\rightarrow\ln\mathbb{E}_\mathcal{U} [{\rm Tr}[\tilde{\rho}_R(o_S)]^2]\,,\\
        \mathbb{E}_{\cal U}[\ln p(o_S)^2]&\rightarrow\ln\mathbb{E}_\mathcal{U}[p(o_S)^2]\,.
    \end{split}
    \label{eq:selfavg-supp}
\end{equation}
In this section, we provide justification in the form of numerical evidence for this self-averaging property.
To do so we define the function
\eq{
{\cal V}[.]=\Big|\frac{\ln\mathbb{E}_\mathcal{U}[.]-\mathbb{E}_\mathcal{U}[\ln [.]]}{\mathbb{E}_\mathcal{U}[\ln [.]]}\Big |\,
}
which is nothing but the relative difference between the quenched and annealed averages. A systematic decay of ${\cal V}[.]$ with $N$ is compelling evidence for the validity of the self-averaging approximation, and to justify both lines of Eq.~\ref{eq:selfavg-supp} we present numerical results for two quantities,
\begin{equation}
    \begin{split}
        \chi_1&=\frac{1}{2^{pN}}\sum_{o_S}{\cal V}[{\rm Tr}[\tilde{\rho}_R(o_S)]^2]\,,\quad\quad\chi_2=\frac{1}{2^{pN}}\sum_{o_S}{\cal V}[p(o_S)^2]\,,\\
    \end{split}
\end{equation}
where the unbiased average over all $o_S$ ensures that our results are not particular to a single choice of $o_S$.
Numerical results for $\chi_1$ and $\chi_2$ (with $p=0.5$ and $\gamma=0.5$ without loss of any generality) are shown in Fig.~\ref{fig:self-avg} where their exponential decay with $N$ provides compelling evidence for the self averaging and hence, for Eq.~\ref{eq:self-avg}.
\end{document}